\documentclass[]{emulateapj}

\usepackage{natbib}
\usepackage{amsmath}
\usepackage{txfonts}
\usepackage{gensymb}
\usepackage{mathrsfs}
\usepackage{epstopdf}
\usepackage{txfonts}
\usepackage{graphicx,color}

\newcommand{\rp}{R_{\rm p}}
\newcommand{\rH}{r_{\rm H}}

\newcommand{\rs}{r_{\rm s}}

\newcommand{\Mp}{M_{\rm p}}
\newcommand{\mj}{M_{\rm J}}
\newcommand{\vz}{v_{\rm z}}
\newcommand{\Sgap}{\Sigma_{\rm gap}}
\newcommand{\eqnref}[1]{Equation \ref{#1}}
\newcommand{\secref}[1]{Section \ref{#1}}
\newcommand{\figref}[1]{Figure \ref{#1}}

\shorttitle{Single Planet Gaps in 3D}
\shortauthors{Fung \& Chiang}

\begin{document}

\title{Gap Opening in 3D: Single Planet Gaps}
\author{Jeffrey Fung\altaffilmark{1} \& Eugene Chiang}
\affil{Department of Astronomy, University of California at Berkeley, Campbell Hall, Berkeley, CA 94720-3411}
\altaffiltext{1}{NSERC Fellow}

\email{email: jeffrey.fung@berkeley.edu}

\begin{abstract}
Giant planets can clear deep gaps when embedded
in 2D (razor-thin) viscous circumstellar disks.
We show by direct simulation
that giant planets are just as capable of carving
out gaps in 3D. Surface density maps
are similar between 2D and 3D, even in detail.
In particular, the scaling $\Sigma_{\rm gap} \propto q^{-2}$
of gap surface density
with planet mass, derived from a global ``zero-dimensional''
balance of Lindblad and viscous torques,
applies equally well to results obtained at higher dimensions.
Our 3D simulations reveal extensive, near-sonic, meridional flows
both inside and outside the gaps; these large-scale circulations
might bear on disk compositional gradients, in dust or other
chemical species. At high planet mass, gap edges are
mildly Rayleigh unstable and intermittently shed streams of material
into the gap --- less so in 3D than in 2D.

\end{abstract}

\keywords{accretion, accretion disks --- methods: numerical --- planets and satellites: formation --- protoplanetary disks --- planet-disk interactions --- circumstellar matter --- stars: variables: T Tauri, Herbig Ae/Be}

\section{Introduction}
\label{sec:intro}
Giant planets are expected to open gaps in their
natal circumstellar disks (for a review,
see \citealt{Kley12}).
Gaps and cavities have been observed
in ``transitional'' disks such as LkCa 15 \citep{Kraus12,Sallum15}, TW Hya \citep{Debes13,Akiyama15,Nomura16}, and HD 169142 \citep{Quanz13,Reggiani14}, in addition to the young protoplanetary disk HL Tau \citep{ALMA_HLTau}. Whether we can say with confidence that these gaps are planetary in origin relies on our understanding of the gap-opening mechanism.

Theoretical studies supported by numerical simulations have
investigated how the depths and widths of planetary gaps relate to
planet and disk properties
\citep{Crida06,Duffell13,Fung14,Kanagawa15a,Kanagawa16}.
It has been established from 2D numerical simulations that the surface density within the gap, $\Sgap$, follows a power-law scaling relation with planet mass, disk aspect ratio, and disk viscosity. This relation can be
derived from a simple ``zero-dimensional'' (0D)
model, where the planet's global
disk-integrated Lindblad torque balances the disk viscous torque
(\citealt{Fung14}, hereafter FSC14; \citealt{Kanagawa15a};
\citealt{Duffell15}). 
Whether these results carry over to a realistic 3D disk is not
obvious. The Lindblad torque is most strongly exerted in the disk midplane
where the planet resides, while the viscous torque applies throughout the
disk.
\citet{Kley01} and \citet{Bate03} reported good agreement
between 2D and 3D simulations for a restrictive set of parameters
and simulation times
(see their Figures 7 and 2, respectively).\footnote{These
papers and others of that era argue that
gaps open when a planet's Hill sphere radius exceeds the disk's
scale height. This criterion is incorrect; even low-mass planets
can open gaps, depending on the disk viscosity (\citealt{Duffell13};
FSC14; \citealt{Kanagawa15a}; \citealt{Kanagawa16};
see also \citealt{Goldreich80}, their equation 103).}
Morbidelli et al.~(\citeyear{Morbidelli14}; see also
\citealt{Szulagyi14})
reported a meridional circulation
inside a planetary gap in their 3D simulations.

Because of the difference in dimensionality, 3D flow patterns cannot be
derived from 2D. Flow dynamics specific to 3D can affect disk
and planet evolution. For
instance, 3D effects may modify the amount of gas that flows around the
planet and across the gap, affecting how the planet accretes mass
\citep[e.g.,][]{Ormel15b} and migrates \citep[e.g.,][]{Benitez15}.
Also, 3D flow patterns around gap edges may impact
dust filtration and trapping at gap edges, processes so far
studied mainly in 2D \citep[e.g.,][]{Paardekooper06a,Rice06,Zhu12}. 

A technical challenge with 3D planetary gap studies is the long
computational timescale involved.
As shown by FSC14, the time it takes for a gap to
come to steady state scales with the disk viscous time, and can
be as long as $\sim$10$^4$ planetary orbits when $\alpha$, the
Shakura-Sunyaev viscosity parameter \citep{alpha}, is 0.001.
This long time, together with the generally more costly
nature of 3D simulations, has made it difficult to simulate 3D gaps.
Fortunately, these obstacles are gradually being removed by
technological advances. In this paper, we perform a systematic
parameter space study that reveals the steady-state density and flow
structure of 3D gaps, which we will compare to 2D results. 

\secref{sec:numerics} details our numerical method and simulation parameters. \secref{sec:gaps} compares 3D gap results with
their 2D counterparts. \secref{sec:flow} details the 3D flow field around gaps, and \secref{sec:Rayleigh} investigates the origin of gap unsteadiness. \secref{sec:conclude} concludes and discusses future work.

\section{Numerical Method}
\label{sec:numerics}
We use the graphics processing unit (GPU) accelerated hydrodynamics
code \texttt{PEnGUIn} \citep{MyThesis} to simulate planetary gaps. It
is a 3D Lagrangian-remap shock-capturing code that uses the piecewise
parabolic method \citep{PPM}. A 2D version of this code was used by
FSC14. A 3D version was implemented by \citet{Fung15} to simulate
the flow field around a low-mass planet, using a setup similar
to the one employed here. In this section, we recapitulate
some of the main features of \texttt{PEnGUIn}, which we will
run in both 2D and 3D.

Denoting by $\{r,\phi,\theta\}$ the radial, azimuthal, and polar
coordinates, and $R=r\sin\theta$ the cylindrical radius, we write
the Lagrangian
continuity and momentum equations solved by \texttt{PEnGUIn} as:
\begin{align}
\label{eqn:cont_eqn}
\frac{D\rho}{Dt} &= -\rho\left(\nabla\cdot\mathbf{v}\right) \,,\\
\label{eqn:moment_eqn}
\frac{D\mathbf{v}}{Dt} &= -\frac{1}{\rho}\nabla p + \frac{1}{\rho}\nabla\cdot\mathbb{T}  - \nabla \Phi \,,
\end{align}
where $\rho$ is the gas density, $\mathbf{v}$ the velocity field, $p$ the gas pressure, $\mathbb{T}$ the Newtonian stress tensor, and $\Phi$ the combined gravitational potential of the star and the planet. The simulations are performed in the corotating frame of the planet, with the Coriolis force absorbed into the conservative form of \eqnref{eqn:moment_eqn} as suggested by \citet{Kley98}. We adopt a locally isothermal equation of state, such that $p= c^2 \rho$, where $c$ is the specified sound speed of the gas. The stress tensor $\mathbb{T}$ is proportional to the kinematic viscosity $\nu$, which we parameterize using the $\alpha$-prescription of \citet{alpha}, such that $\nu=\alpha c h$, where $h$ is the disk scale height. We fix $\alpha=0.001$.

The simulations are performed in 
a frame centered on the star,
so for a planet on a fixed, circular orbit in the disk midplane,
$\Phi$ equals:
\begin{equation}\label{eqn:potential}
\Phi = -\frac{GM}{1+q}\left[\frac{1}{r} + \frac{q}{\sqrt{r^2 + \rp^2 - 2R\rp\cos{\phi'} + \rs^2}} - \frac{qR\cos{\phi'}}{\rp^2}\right] \,,
\end{equation}
where $G$ is the gravitational constant, $M=M_* + \Mp$ the total mass of the star and the planet, $q = \Mp / M_*$ the planet-to-star mass ratio, $\rp$ the semi-major axis of the planet's orbit, $\rs$ the smoothing length of the planet's potential, and $\phi' = \phi-\phi_{\rm p}$ denotes the azimuthal separation from the planet. We set $GM=1$ and $\rp=1$, so that the Keplerian velocity and frequency $v_{\rm k} = \sqrt{GM/r}$ and $\Omega_{\rm k} = \sqrt{GM/r^3}$ both equal 1 at the planet's orbit. We also define $\Omega_{\rm p}$ as the planet's orbital frequency. The third term in the bracket is the indirect potential. The 2D version of \eqnref{eqn:potential} is obtained by substituting the spherical $r$ with the cylindrical $R$. Note that \eqnref{eqn:potential} technically differs from the potential used in FSC14, as \texttt{PEnGUIn} simulations in FSC14 were performed in the barycentric frame. This difference in reference frame should not have a significant effect on our results, as was already tested by FSC14.

One qualitative difference between 2D and 3D setups is in the smoothing length $\rs$. In 2D, $\rs$ is chosen to mimic the vertically averaged gravitational force of the planet, and therefore the appropriate choice should be of order the thickness of the disk, $h = c/\Omega_{\rm k}$ \citep{Muller12}. In 3D, $\rs$ should be set to a value sufficiently small to not interfere with the large-scale gap dynamics of interest here (in contradistinction to the smaller-scale circumplanetary flow). Based on these considerations, we set $\rs = 0.5~h_0$ in our 2D simulations, where $h_0$ is the disk scale height evaluated at the planet's location, and $\rs = 0.1~\rH$ in 3D, where $\rH = (q/3)^{1/3}\rp$ is the planet's Hill radius. 

\subsection{Initial and boundary conditions}
\label{sec:initial}

The initial disk profile assumes the following surface density and sound speed profiles:
\begin{equation}\label{eqn:initial_sigma}
\Sigma =  \Sigma_0 \left(\frac{R}{\rp} \right)^{-\frac{1}{2}}\, ,
\end{equation}
\begin{equation}\label{eqn:initial_c}
c =  c_0 \left(\frac{R}{\rp} \right)^{-\frac{1}{2}}\, ,
\end{equation}
where we set $\Sigma_0 = 1$,\footnote{Since we do not consider the self-gravity of the disk, this normalization has no impact on our results.} and choose $c_0 = 0.05$. The disk scale height $h$ is therefore
$c/\Omega_{\rm k} = 0.05R$ with $\Omega_{\rm k}$ evaluated at the midplane; in other words, the disk aspect ratio $h/R=0.05$ is constant. 
This sound speed profile is fixed in time, corresponding to an irradiated disk that has a cooling time much shorter than the dynamical time. In 3D hydrostatic equilibrium, the initial density structure is:
\begin{equation}\label{eqn:initial_rho}
\rho = \rho_0 \left(\frac{R}{\rp}\right)^{-\frac{3}{2}} \exp\left(\frac{GM}{c^2} \left[\frac{1}{r}-\frac{1}{R}\right]\right) \,,
\end{equation}
where $\rho_0 = \Sigma_0/\sqrt{2\pi h_0^2}$ is the initial gas density at the location of the planet.

The initial velocity field assumes hydrostatic equilibrium, taking into account gas pressure but not viscosity. As a result, the initial disk has an orbital frequency of
\begin{equation}\label{eqn:omega}
\Omega = \sqrt{\Omega_{\rm k}^2 + \frac{1}{r}\frac{\partial p}{\partial r}} 
\end{equation}
and zero radial and polar velocity. 
Given our isothermal equation of state and
sound speed profile, our disk model is subject to the vertical shear
instability \citep[e.g.,][]{Urpin98,Lin15}.
However, our disk viscosity of $\nu= 2.5\times10^{-6}$
is sufficient to prevent the instability from
growing, as shown by \citet{Nelson13}.

Both the inner and outer radial boundaries are fixed at their initial
values.
In the polar direction, we simulate only the upper half of the disk to
reduce computational cost.
We use reflecting boundaries both at the top, to prevent mass from 
entering or leaving the domain, and at the midplane.
Imposing reflection
symmetry across the midplane (polar velocity $v_\theta = 0$)
means certain dynamics are not captured
by our simulation, such as 
flows that cross the midplane, and bending waves near
the midplane. It remains to be determined whether these restrictions
affect the results of our study.

\subsection{Resolution}\label{sec:resolve}
Our simulation domain extends radially from $0.4$ to 2 $\rp$, and spans the full $2\pi$ in azimuth. The polar angle in our 3D simulations spans $0$ to $12$ degrees, equivalent to a vertical extent of 4 scale heights.

We use 270 logarithmically spaced cells in the radial direction, 810 uniformly spaced cells in the azimuthal direction, and 45 uniformly spaced cells in the polar direction. This resolution translates to about 8, 6, and 11 cells per scale height in each of the $\{r,\phi,\theta\}$ directions, respectively. This choice is motivated by Figure 2 of FSC14.

\subsection{Metrics}\label{sec:metric}
As in FSC14, we define the surface density in the gap, $\Sgap$, as a spatial average over the annulus spanning $R= \rp -\Delta$ to $\rp + \Delta$ with $\Delta \equiv 2~\rH$, excised from $\phi = \phi_{\rm p} -\Delta/\rp$ to $\phi_{\rm p} + \Delta/\rp$. Throughout this paper, we will also show velocity plots and maps that are temporal and/or spatial averages. In all instances, velocities are averages weighted by gas density.

\section{Results}
\label{sec:results}
\subsection{Gap depth and shape}
\label{sec:gaps}

\begin{figure}[]
\includegraphics[width=0.99\columnwidth]{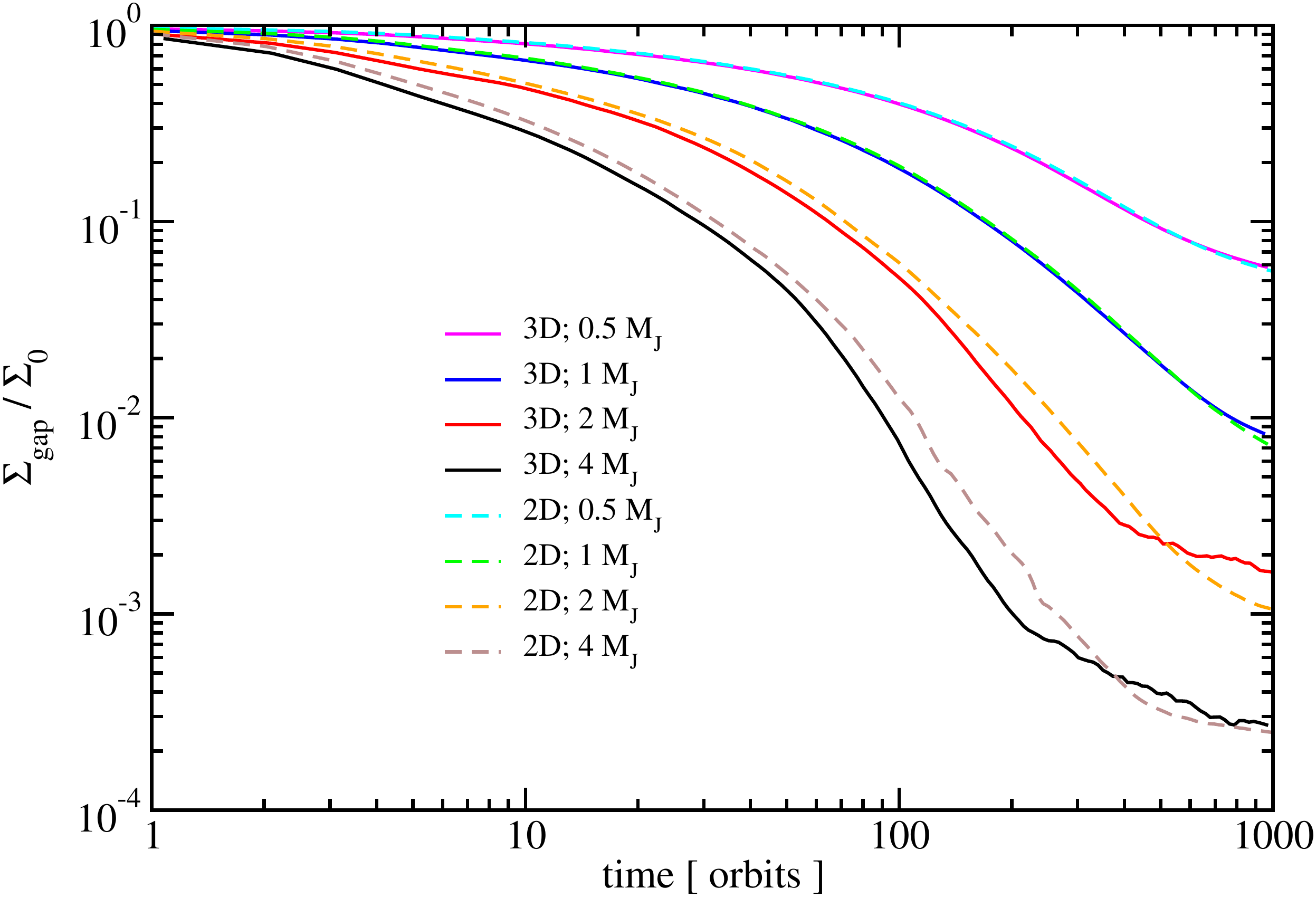}
\caption{$\Sgap$ versus time for all our simulations. Solid (dashed)
lines plot our 3D (2D) results. Comparing our 2D
results to those of FSC14 (their Table 1 and Figure 5),
we find that our measurements of $\Sgap$ have converged
to within a factor of 2 of theirs after 1000 orbits.
The agreement between our 2D and 3D results is close,
particularly for $0.5~\mj$ and $1~\mj$.}
\label{fig:time}
\end{figure}

\begin{figure}[]
\includegraphics[width=0.99\columnwidth]{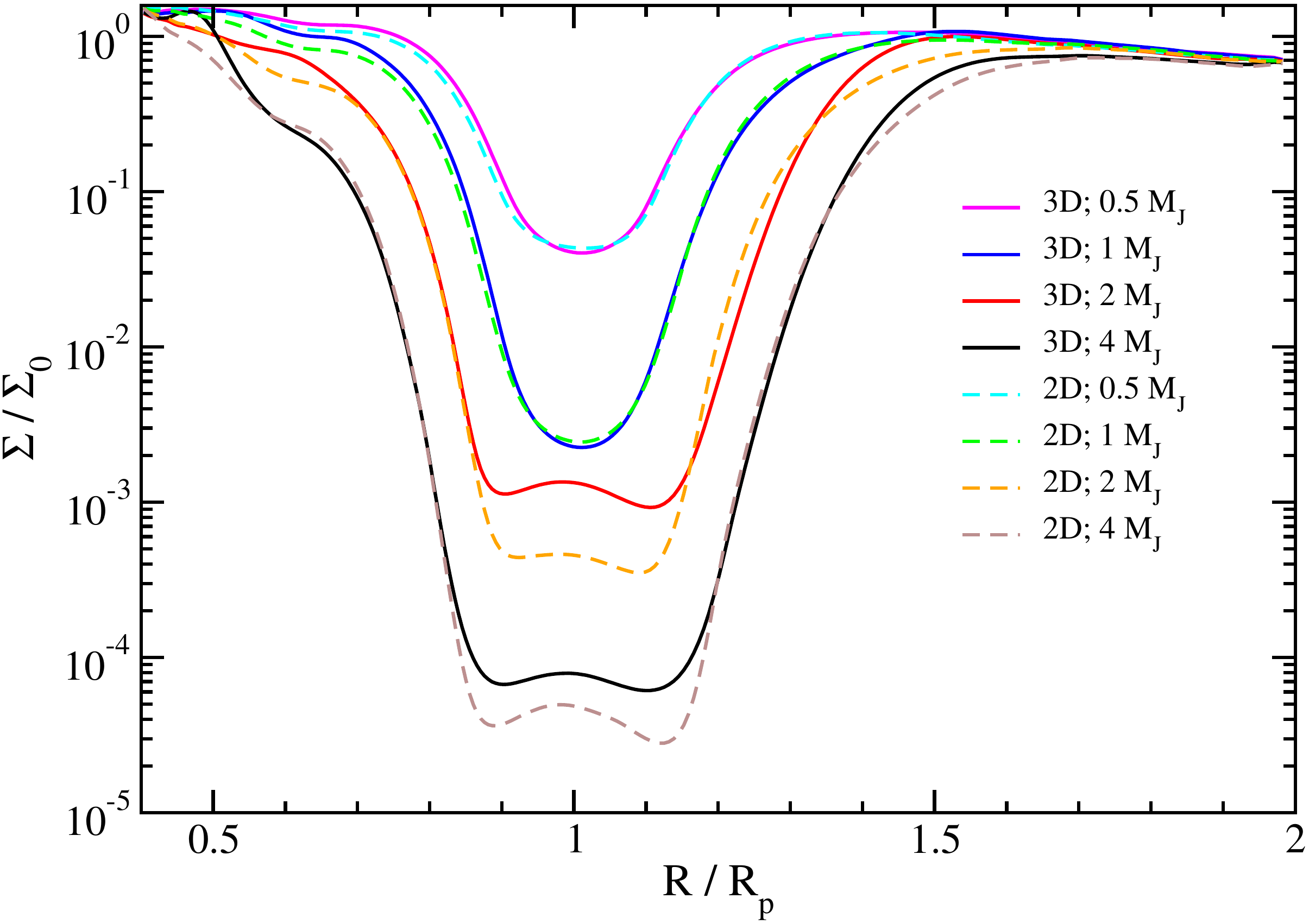}
\caption{Surface density profiles, averaged
azimuthally except for a small area around the planet
($\{\phi_{\rm p}-\Delta/\rp, \phi_{\rm p}+\Delta/\rp\}$),
and averaged in time from 1000 to 1010 planetary orbits.
The slight density enhancement at $R=R_{\rm p}$ when 
$\Mp\geq2\mj$
is due to gas concentrating about the 
triangular (i.e., Trojan) Lagrange points.}
\label{fig:shape}
\end{figure}

\begin{figure*}[]
\includegraphics[width=1.99\columnwidth]{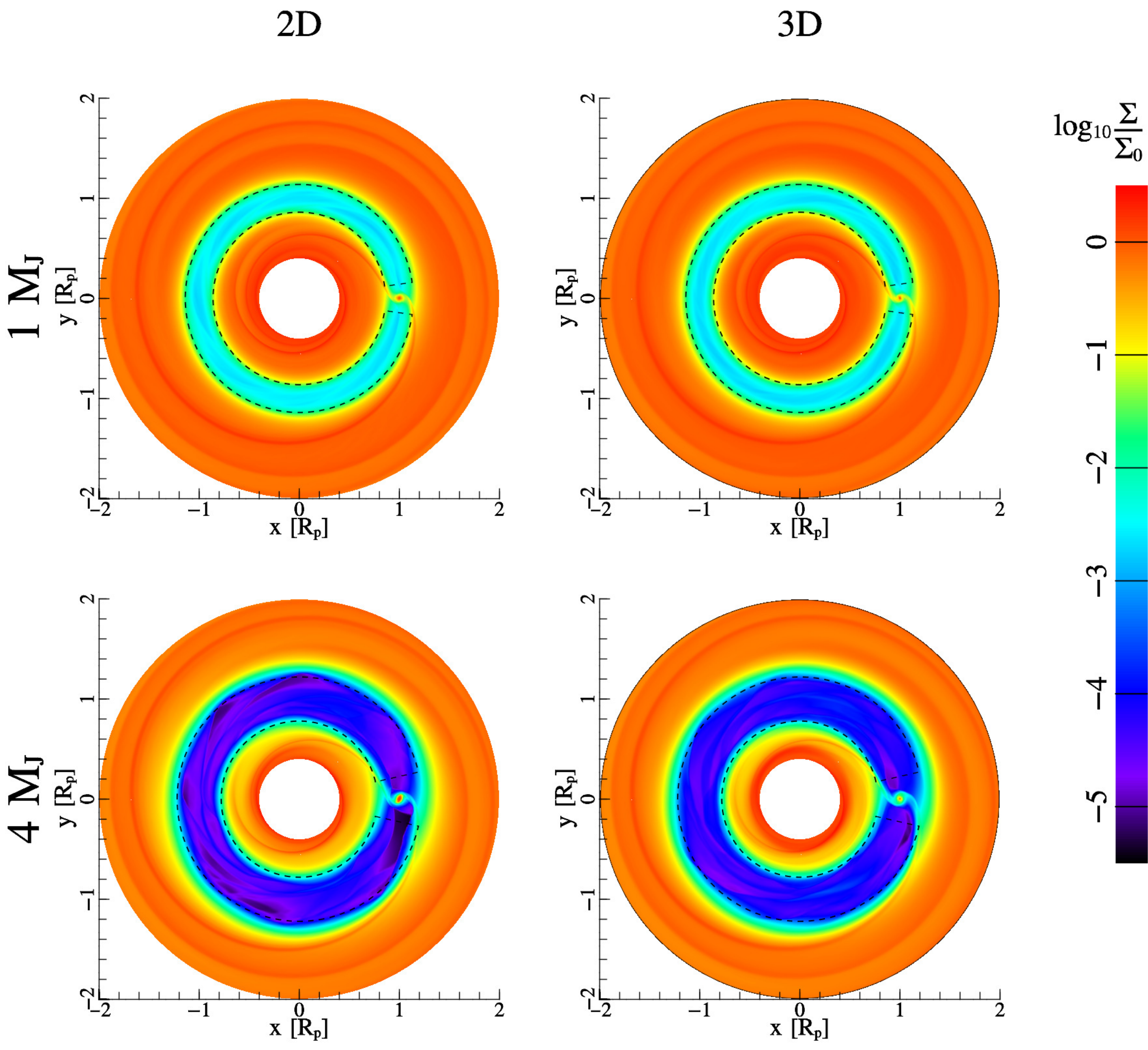}
\caption{Surface density snapshots for our
$1\mj$ and $4\mj$ simulations, taken at 1000 planetary orbits.
Gaps in 3D (right) and 2D (left) have similar depths and widths.
Black solid lines enclose the region where $\Sgap$ is evaluated (see
\secref{sec:metric} and \figref{fig:time}).
Gap streamers are evident in all snapshots, particularly in 2D and for $4\mj$,
and are likely caused by the Rayleigh instability at gap
edges (see \secref{sec:Rayleigh}).}
\label{fig:disks}
\end{figure*}

We perform four 3D simulations with $q$ varying between 0.0005 and 0.004 
(0.5 to 4 $\mj$).
For our parameters, the planet's Hill radius $\rH$
ranges from $1.1$ to $2.2$ times the local disk scale height $h_0$.
We also run four 2D simulations using the same parameters for comparison;
these are nearly exact replicates of the simulations from FSC14. 
\figref{fig:time} plots the gap depth contrast $\Sgap/\Sigma_0$ versus
time for our 2D and 3D simulations.
Because of the costly nature of 3D simulations, we were only able to
simulate up to $10^3$ orbits. By comparison,
FSC14 ran up to $\sim$10$^4$ orbits to achieve steady state. Comparing
our \figref{fig:time} with Figure 5 of FSC14 indicates that
our results for $\Sgap/\Sigma_0$
have converged to within a factor of 2 of theirs.
We consider such convergence sufficient to identify differences
between 2D and 3D gaps. Our results show some time variability,
particularly for high-mass runs, that we will investigate in
\secref{sec:Rayleigh}.

Figure \ref{fig:time} shows clearly that $\Sgap$ behaves in 3D
similarly to how it does at lower dimensionality.
In particular, the final values of $\Sgap$ in 3D
scale as $q^{-2}$, as predicted by the 0D models of FSC14 and
\citet{Kanagawa15a}. The agreement between 2D and 3D simulations is
astonishingly close for $\Mp\leq\mj$. For $\Mp=2\mj$ and $4\mj$, $\Sgap$
decreases initially more rapidly in 3D than in 2D, but ultimately
reaches a slightly larger value (about $\sim$80\% larger in
the case $\Mp=2\mj$).

For 3D simulations, to verify that our fiducial choice for
smoothing length $\rs = 0.1 \rH$
is sufficiently small, we perform a test case 
at $\Mp = 4\mj$ with $\rs = 0.05 \rH$,
and find that $\Sgap$ changes negligibly. For 2D simulations,
the correct choice of $\rs$ is less clear; although
any value of order $h_0$ is suitable in principle \citep{Muller12},
the value of $\Sgap$ is sensitive to $\rs$.
For example, we find that using $\rs = 1~h_0$ instead of
our fiducial $0.5~h_0$ results in a $\sim$40\% larger $\Sgap$.
In this regard, we should not overinterpret 
the exceptionally close agreement between 2D and 3D results.

\figref{fig:shape} plots the
azimuthally averaged surface density profiles, excluding a small
$\pm \Delta/R_{\rm p}$ azimuthal
range around the planet. In terms of both gap width
and the sharpness of gap edges, the 3D gaps are remarkably similar
to 2D gaps. 
The resemblance is confirmed in face-on views of surface
density, as presented in \figref{fig:disks}. The 2D gaps show
more distinct streamers within the gaps.
These filaments appear related to the Rayleigh instability
operating at gap edges, as we discuss in \secref{sec:Rayleigh}.

\subsection{3D Flow Topology}
\label{sec:flow}

\begin{figure*}[]
\includegraphics[width=1.99\columnwidth]{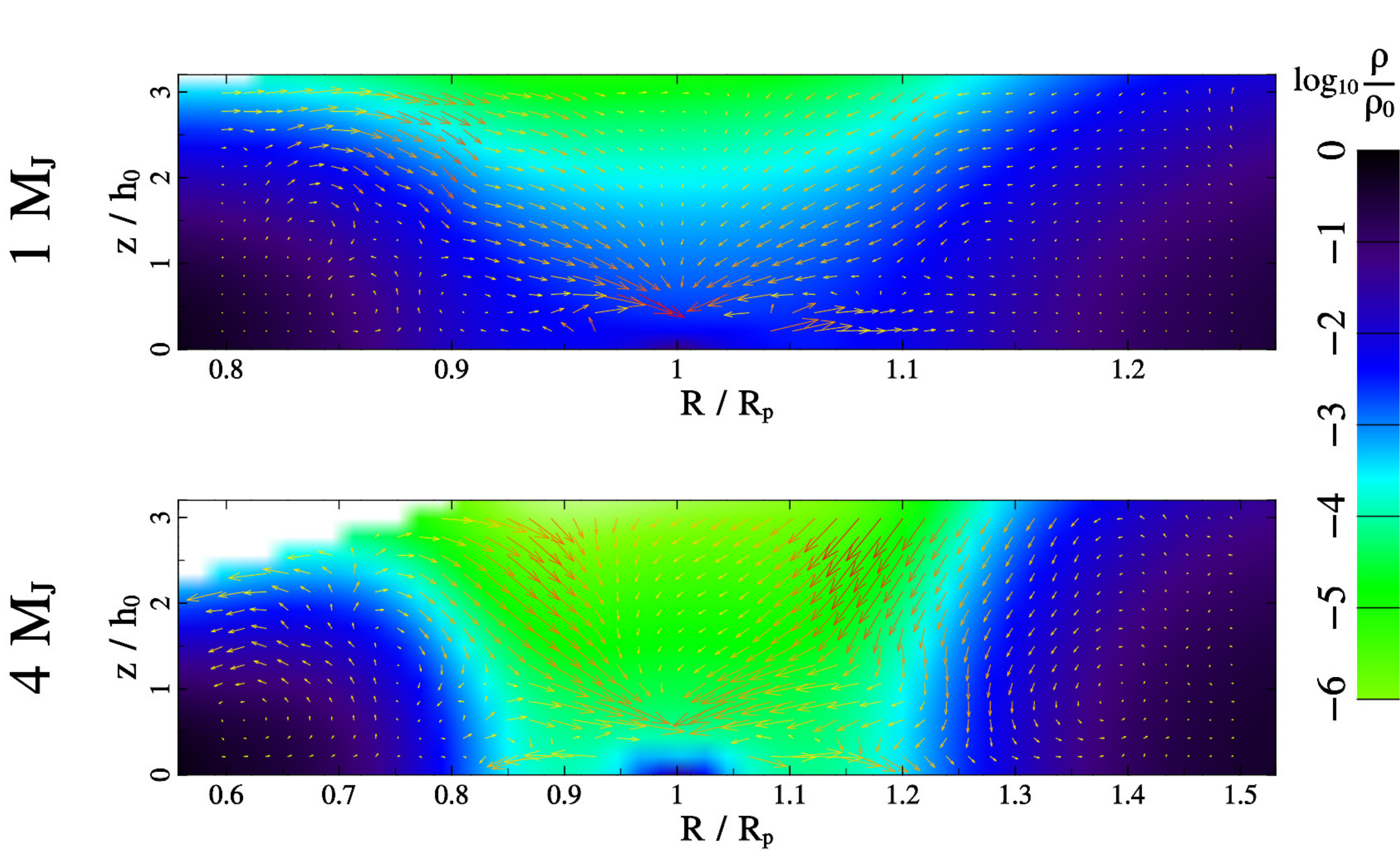}
\caption{Meridional flow fields obtained at the end of our
$1~\mj$ and $4~\mj$ simulations. Arrows represent velocity
vectors, and arrow lengths represent speeds. The longest arrows correspond 
to $0.2~c_0$ for the top panel, and $0.4~c_0$ for the bottom.
Background color represents gas density. 
Both density and velocity are azimuthally averaged over the entire
$2\pi$, and time averaged from 1000 to 1010 planetary orbits.
Velocity vectors within $0.5~\rH$ of the planet are omitted for clarity.
We note that the bottommost row of arrows represent velocities
near but not exactly at the midplane. Because of the reflective
boundary we impose at the midplane, the polar velocity at $\theta=\pi/2$ 
(the midplane) is strictly
zero.}
\label{fig:flow_speed}
\end{figure*}

\begin{figure*}[]
\includegraphics[width=1.99\columnwidth]{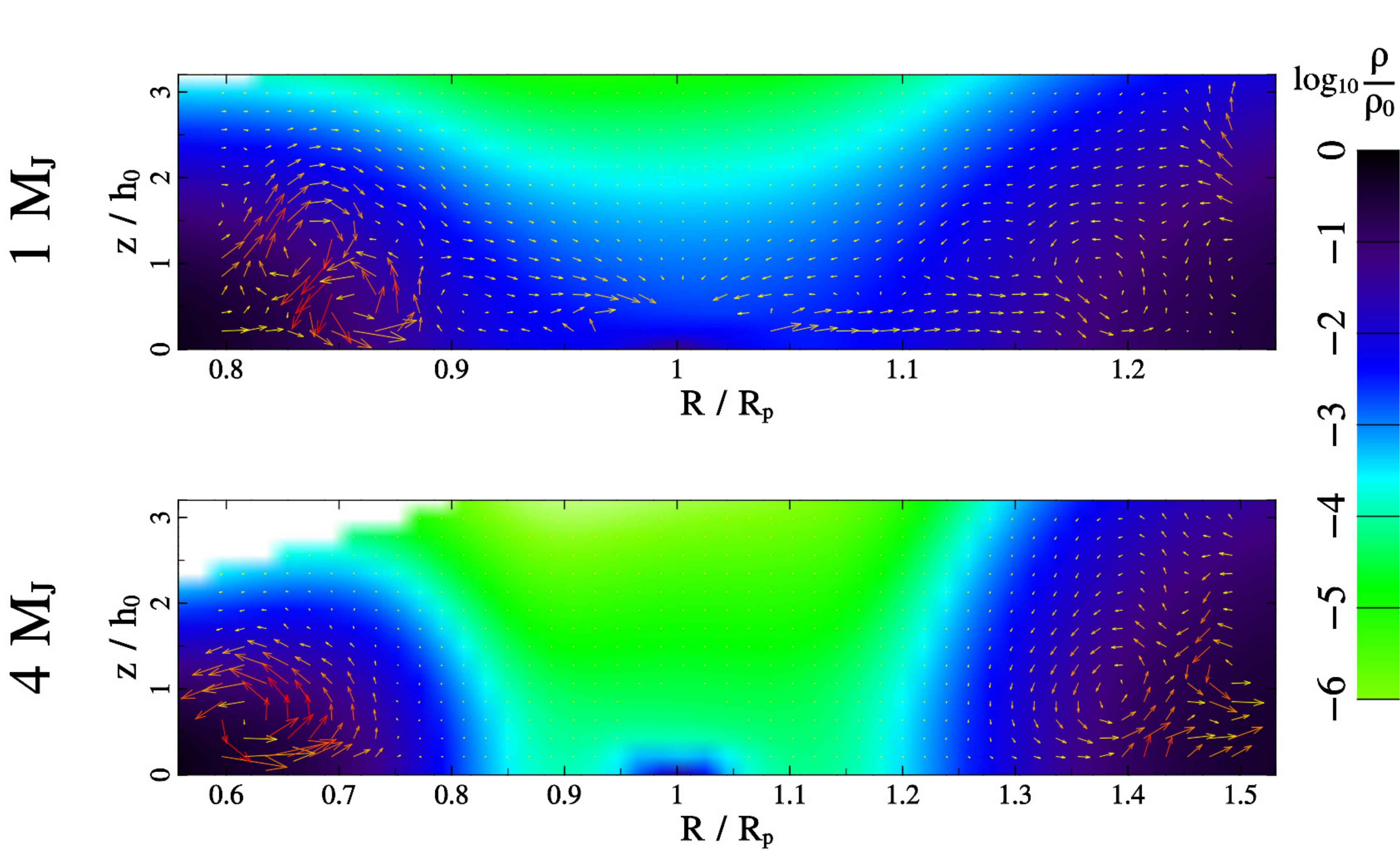}
\caption{Similar to \figref{fig:flow_speed}, but with arrows representing
meridional momentum (averaged over azimuth and time;
the longest arrows correspond to $0.0014~\rho_0c_0$ for the top panel, and 
$0.0073~\rho_0c_0$ for the bottom).
Plotting the momentum highlights the eddies just outside the gap edges;
plotting the velocity as in \figref{fig:flow_speed} highlights the flow
inside the gap.}
\label{fig:flow_momt}
\end{figure*}

\begin{figure*}[]
\includegraphics[width=1.99\columnwidth]{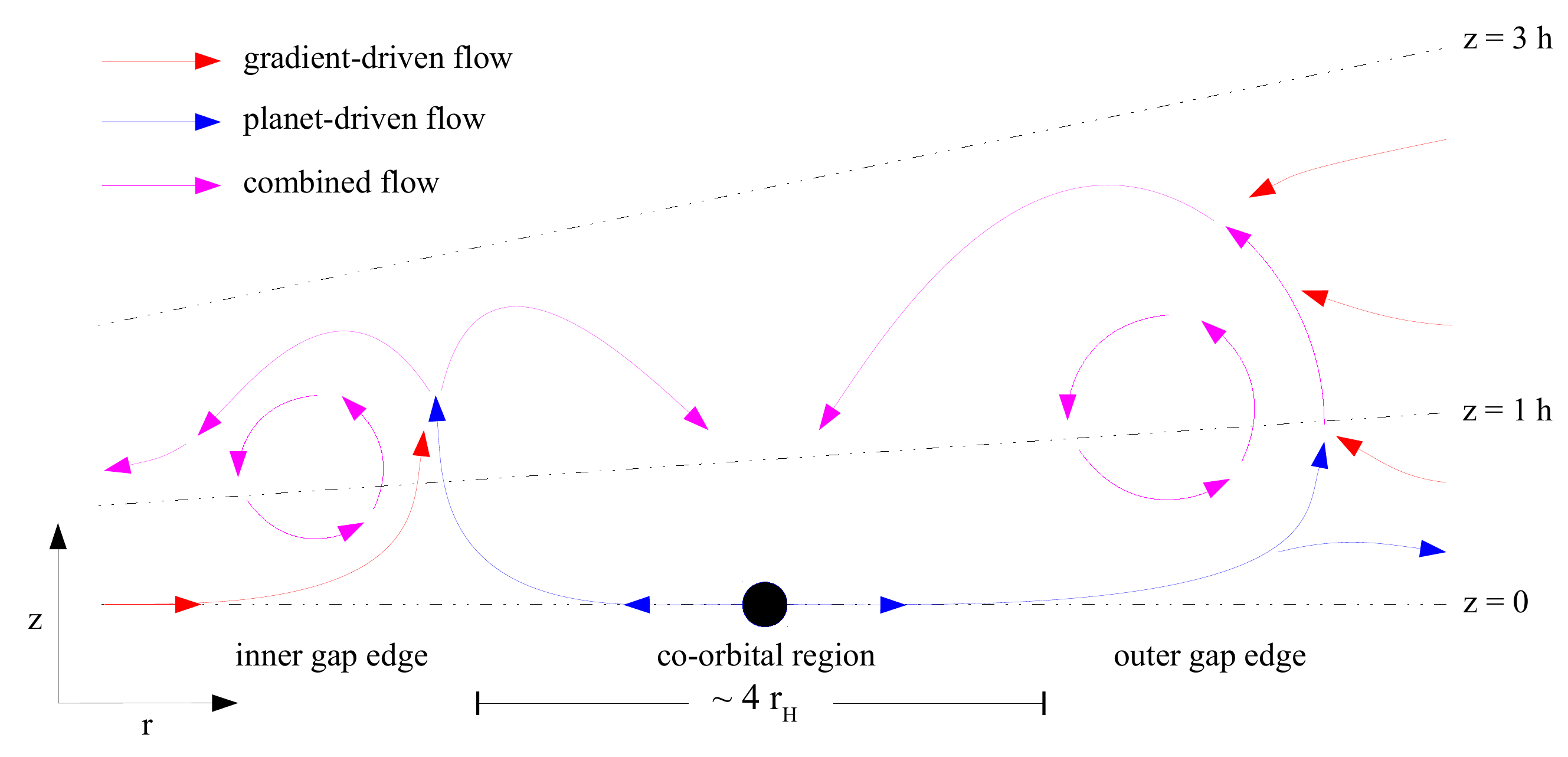}
\caption{A simplified schematic of the meridional flow field,
abstracted from Figures \ref{fig:flow_speed} and \ref{fig:flow_momt}.
The ``planet-driven flow'' (represented by blue arrows) is
generated by the planet's repulsive Lindblad torques.
The ``gradient-driven flow'' (red arrows) is driven by viscous torques
acting to diffuse material from the gap edges into the gap. 
The collision of these two flows near the midplane increases
the density and pressure there relative to vertical hydrostatic equilibrium,
forcing the merged flow (magenta arrows) upward and
driving a meridional circulation.
The eddies at gap edges are strongest near the planet
and persist downstream of its location
for $\sim$1--2 rad in azimuth.
}
\label{fig:schm}
\end{figure*}

\begin{figure*}[]
\includegraphics[width=1.99\columnwidth]{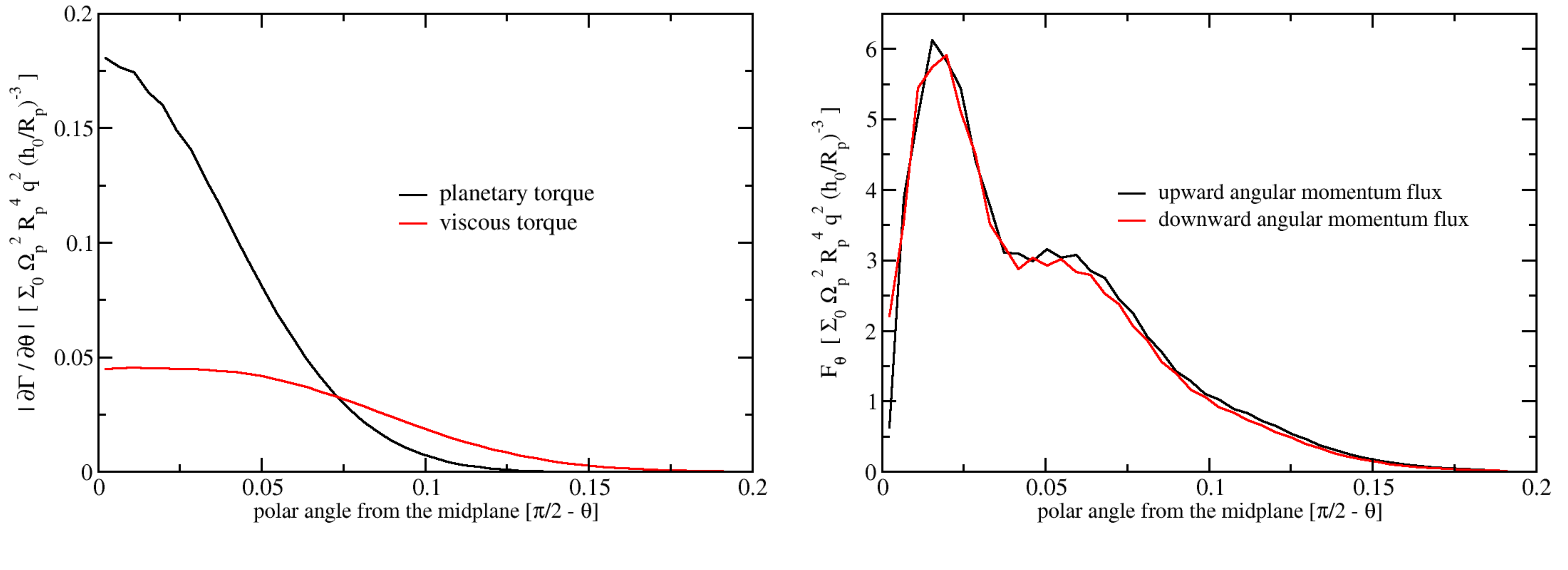}
\caption{Left: absolute values of the planetary (black) and viscous (red) torques, differentiated along the polar angle. Right:  angular momentum flux in the polar direction, with the upward (away from the midplane) component  plotted in black, and the downward (toward the midplane) component in red. The values are obtained from the $1\mj$ simulation, integrated radially from $r=1.1$ to $1.8$ and azimuthally over $2\pi$, and time averaged from 1000 to 1010 orbits. The torque exerted on the disk by the planet is positive, while the viscous torque is negative. The left panel shows that at higher altitudes, the viscous torque overtakes the planetary torque, changing the sign of the net torque. 
The right panel shows that at a given altitude ($\theta$), angular momentum is carried both upward (at certain locations in $r$ and $\phi$) and downward 
(at other locations), with the two fluxes in close but not perfect balance --- their difference is comparable
in magnitude to the planetary and viscous torques
plotted on the left, supporting our interpretation
that torque balance is achieved by vertical
transport.}
\label{fig:tor_z}
\end{figure*}

\begin{figure}[]
\includegraphics[width=0.99\columnwidth]{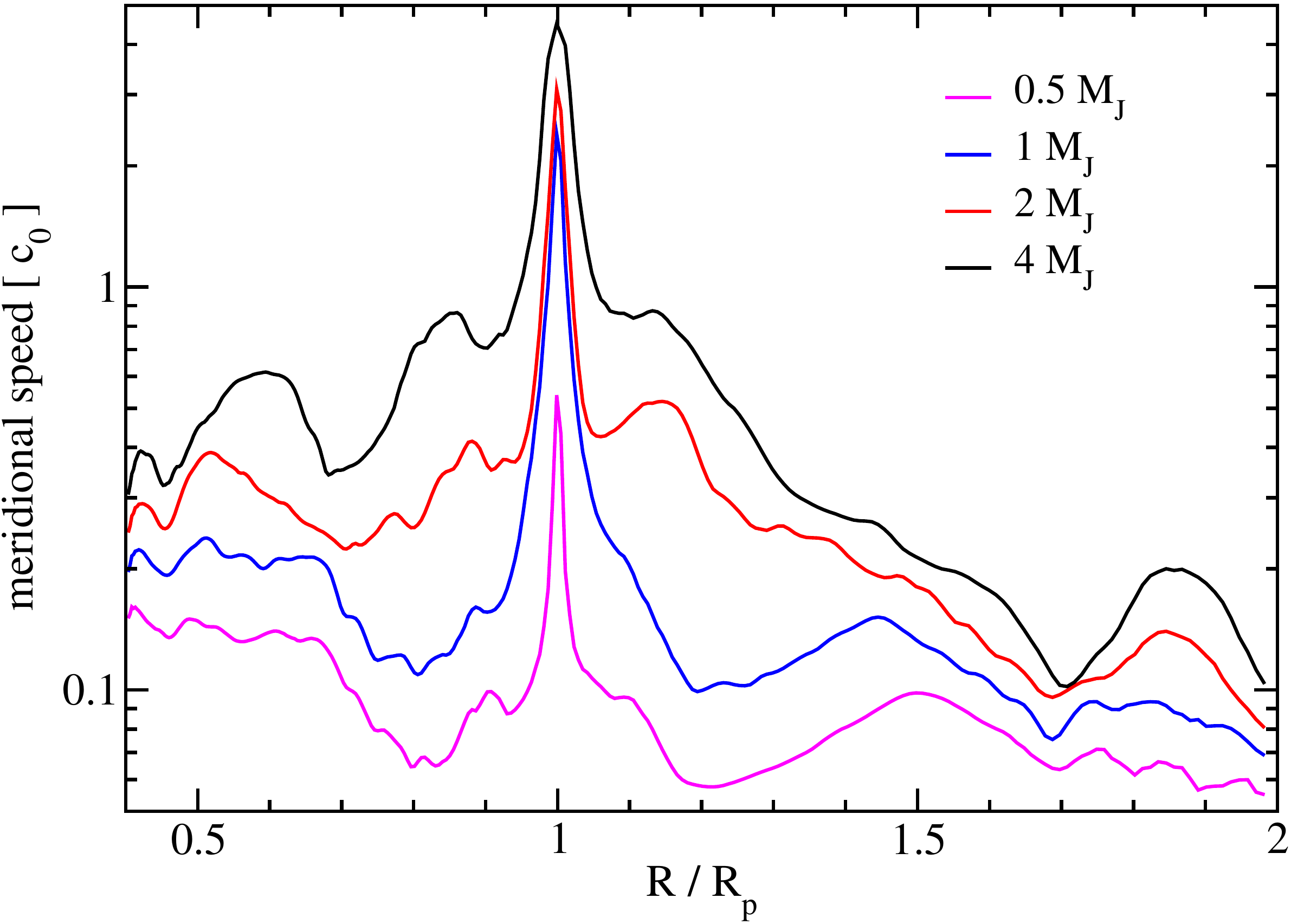}
\caption{Meridional speeds (vertical and radial velocities
added in quadrature, and averaged azimuthally and vertically)
versus cylindrical radius in our 3D simulations. All profiles are
evaluated after 1000 planetary orbits and normalized to $c_0$, the disk sound
speed at the planet's location.
Jupiter-mass planets generate
near-sonic meridional flows over radial distances comparable to their
orbital radii.}
\label{fig:meri_speed}
\end{figure}

\begin{figure*}[]
\includegraphics[width=1.99\columnwidth]{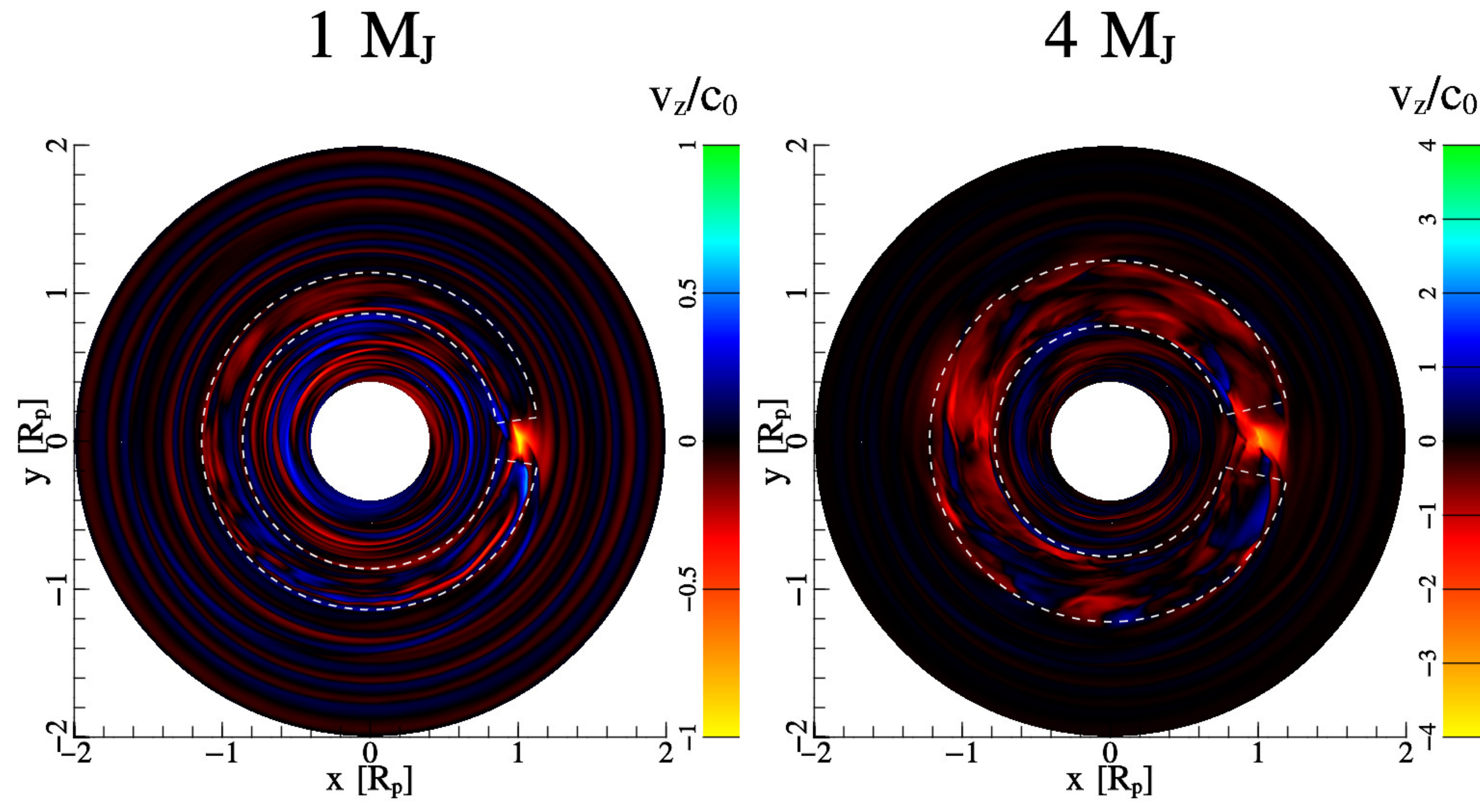}
\caption{Snapshots of the vertically averaged vertical velocity for $1\mj$
(left) and $4\mj$ (right) at 1000 planetary orbits. The vertical average
is taken over the upper half of the disk only; this is our entire
simulation domain, as we assume the bottom half of the disk
is reflection-symmetric about the midplane. The white dashed lines
enclose the region where $\Sgap$ is calculated (see \secref{sec:metric} and
\figref{fig:time}). Blue and green colors represent gas moving with positive
vertical velocities (away from the midplane), while red and yellow indicate
the opposite. Just above the planet, velocities are supersonic and directed
toward it.}
\label{fig:vz_map}
\end{figure*}

The flow pattern in 3D may be expected
to deviate significantly from that in 2D. 
In 3D, not only are vertical velocities allowed,
but radial velocities can be larger since planetary 
and viscous torques do not have to cancel locally; the 
former is mostly exerted in the midplane, while the latter
can be exerted anywhere.

Figures \ref{fig:flow_speed} and \ref{fig:flow_momt} illustrate the
meridional flow in our $1\mj$ and $4\mj$ simulations. In agreement with
the results of \citet{Szulagyi14} and \citet{Morbidelli14}, 
we find that within the gap, there
exists a radial flow directed away from the planet's orbit near the
midplane, and another flow delivering gas back into the gap at higher
altitudes (\figref{fig:flow_speed}). Whereas
\citet{Morbidelli14} describe a closed, localized circulation,
where the gas is trapped in a cyclical flow near the gap,
the flow pattern that we uncover is less localized.
We find vigorous motion outside the gap as well as inside with
multiple eddy-like structures (\figref{fig:flow_momt}).

\figref{fig:schm} is a schematic drawing of the overall meridional
flow field. Outside the gap, steep density
gradients drive a viscous flow into the gap (``gradient-driven''
flow shown in red). At the same time, the planet
drives a fast, denser midplane flow directed away from the gap
(``planet-driven'' flow shown in blue). 
These two flows collide near the gap edges, where more complex flow
patterns emerge. Because midplane disk pressure decreases outward, the
planet-driven flow encounters little resistance moving beyond
the outer gap edge, and is able to reach a radial distance
of $\sim$4 $\rH$ away from the planet's orbit before being deflected.
By comparison, on the opposite, interior half of the gap,
the planet-driven flow
is deflected at $\sim$2 $\rH$ because of the higher gas pressure
characterizing the inner gap edge.
Where the planet-driven flow collides with the 
gradient-driven flow, the local pressure and density are enhanced. 
The build-up of pressure at the midplane drives gas upward
(the ``combined'' flow shown in magenta) which eventually
falls back down toward the midplane.
Meridional eddies are created whose rotation
poles point in the $-\phi$ direction, just outside either gap edge.
The eddies are strongest near the planet and persist
downstream for $\sim$1--2 radians.

The meridional eddies extend as high as our vertical simulation boundaries allow.
Gas is carried all the way from the disk
midplane to a few scale heights above, and back.
This vertical circulation enables the system to 
reach a near-steady state: the angular momentum acquired
by a gas element at low altitude, where the planetary torque dominates,
nearly cancels with the angular momentum acquired at high
altitude, where the viscous torque dominates.
This change in the sign of the net torque is illustrated by the left panel of
\figref{fig:tor_z}, which shows that the planetary torque, integrated 
over a portion of the outer disk for the $1\mj$ case, dominates near 
the midplane, but falls rapidly with altitude and is eventually overtaken 
by the viscous torque. Note that the precise altitude where this sign 
change occurs is radially dependent, because the planetary torque is 
strongest close to the planet, while the viscous torque is strongest 
near the gap edge.
The right panel of \figref{fig:tor_z} plots the angular momentum flux in the 
polar direction, $F_{\theta}$, integrated over the same portion of the outer disk:

\begin{equation}\label{eqn:F_theta}
F_{\theta} = \int^{2\pi}_{0}\int^{1.8}_{1.1} \langle \rho v_{\phi} v_{\theta}  r \sin{\theta}\rangle r ~{\rm d}r ~{\rm d}\phi
\end{equation}
where the brackets indicate a time average
from 1000 to 1010 orbits.
In \figref{fig:tor_z} we separate
$F_{\theta}$ into its upward ($v_{\theta}<0$) and downward ($v_{\theta}>0$) components.
The difference between the upward
and downward fluxes is small but significant: the net
residual is similar in magnitude to the planetary
and viscous torques plotted on the left panel
of \figref{fig:tor_z}. This result is consistent with our interpretation that torque balance is achieved through vertical transport. A more
explicit verification
of the balance is difficult because the flow varies strongly in time and space, and flow patterns
are not strictly closed.

\figref{fig:meri_speed} plots the vertically and azimuthally averaged
meridional speeds of our 3D simulations.
The speeds come within an order of magnitude of the sound speed over
distances comparable to the planet's orbital radius, and are
supersonic closest to the planet (e.g.,
there is a Mach $\sim$ 5 shock deep inside the $4\mj$
planet's Hill sphere).
The planet's influence extends well beyond Hill sphere scales;
for example, the most distant principal Lindblad resonances are located at
$0.63~\rp$ in the inner disk (2:1 resonance), and $1.59~\rp$ in the outer
disk (1:2 resonance).

\figref{fig:vz_map} plots 2D maps of the 
vertical velocity $\vz$, vertically averaged
over the upper half of the disk above the midplane (recall
that our simulations assume reflection symmetry
about the midplane), and gives a sense of the flow's
azimuthal variations.
The fastest downward flows in the gap are concentrated within
$\sim$1 $\rH$ of the planet. 
The velocities are comparable to the sound speed, as befits
gas that has freely fallen a distance $\sim\rH$ over a timescale $\Omega^{-1}$.
Farther away azimuthally,
the flow speed reduces by factors of a few and fluctuates
spatially and temporally. 

Vertical velocities might be used one day to distinguish
planet-opened gaps from other kinds of gaps (e.g., \citealt{Zhang15};
\citealt{Bethune16}), provided velocity signatures above
and below the disk midplane along the observer's line of sight
do not cancel to zero. A net velocity of zero would be obtained
if the disk were exactly reflection-symmetric about
the midplane, optically thin in the line used
to measure Doppler shifts, and viewed face-on. 
Rotational transitions from CO may be optically thick, even
in the low density environment of a gap.
For example, at a temperature of 100 K and a CO/H$_2$ number
abundance of $10^{-4}$, the $J$=4-3 CO line is optically
thick at line center down to $\Sigma \sim 10^{-4}$ g/cm$^2$.

\subsection{Rayleigh instability at gap edges}
\label{sec:Rayleigh}

\begin{figure*}[]
\includegraphics[width=1.99\columnwidth]{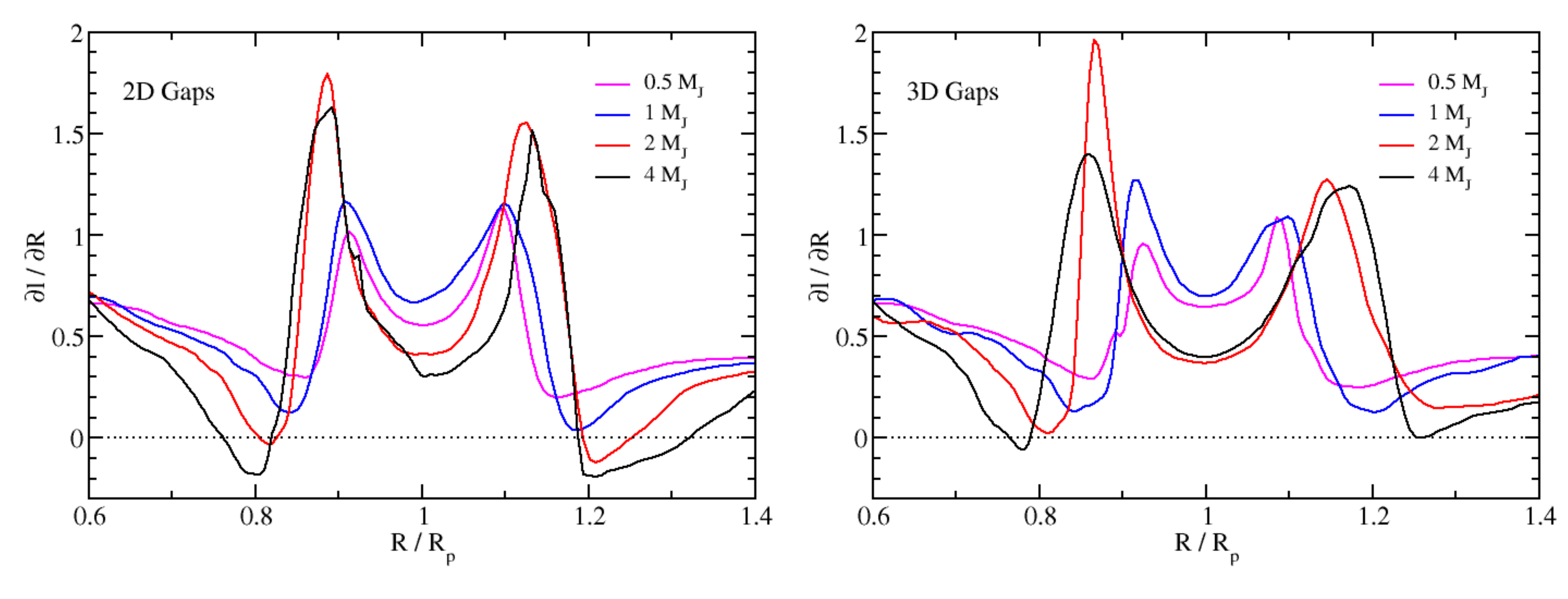}
\caption{Differential angular momentum profiles for all 8 of our simulations, with 2D models on the left and 3D ones on the right. They are azimuthally averaged, excluding the small range around the planet $\phi=\{\phi_{\rm p}-\Delta/\rp, \phi_{\rm p}+\Delta/\rp\}$, and obtained at 1000 planetary orbits. For our 3D models, the profiles are taken at the midplane. Where the profile becomes negative is where the disk becomes Rayleigh unstable. 
The 2D profiles are more unstable than the 3D profiles.}
\label{fig:ang_momt}
\end{figure*}

\begin{figure}[]
\includegraphics[width=0.99\columnwidth]{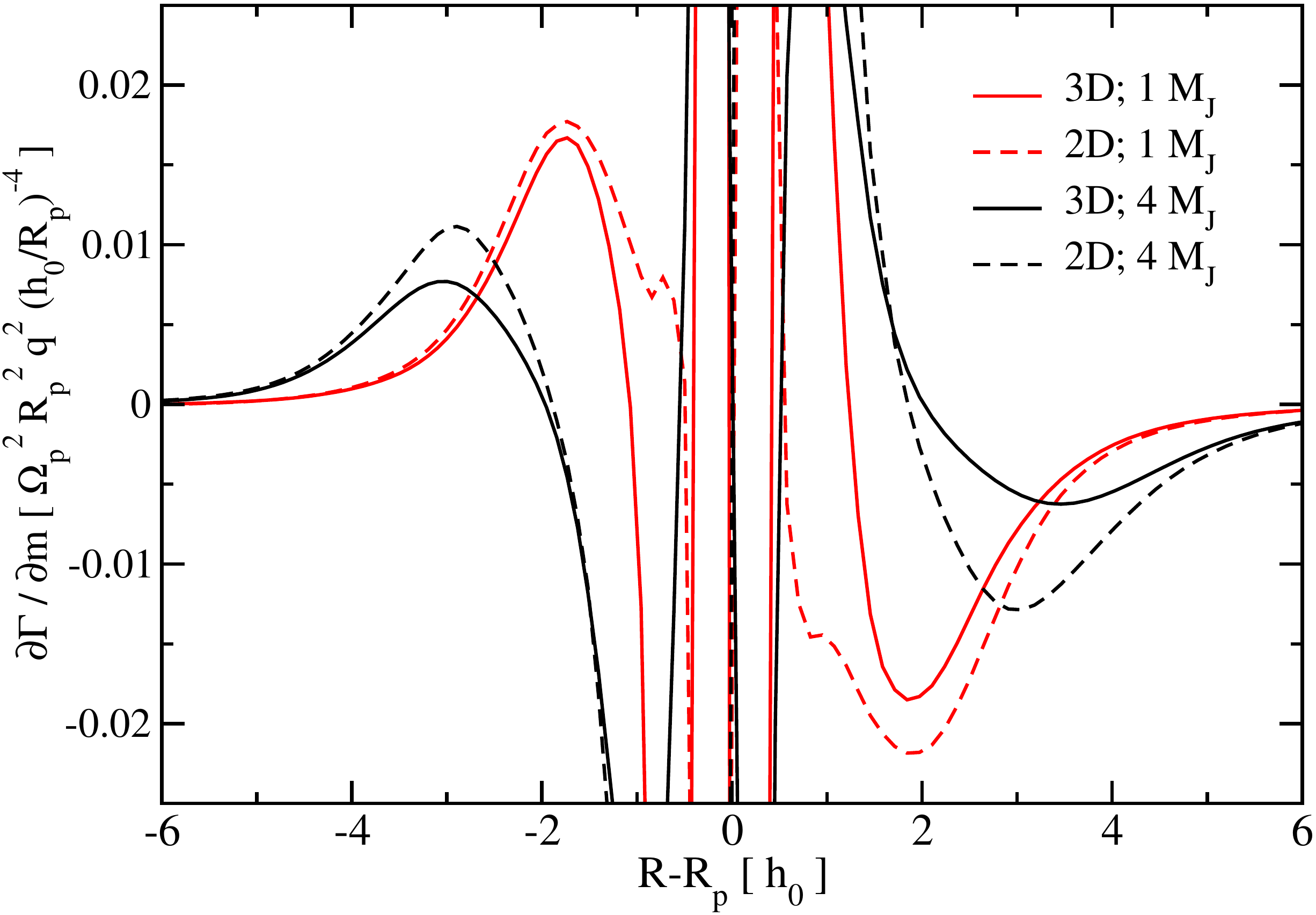}
\caption{Planetary torque density, azimuthally integrated 
and time averaged from 1000 to 1010 planetary orbits,
for our 1 and 4 $\mj$ models.
More than $\sim$1 $h_0$ from the planet, the torque profiles
show qualitatively good
agreement between 2D (dashed) and 3D (solid) calculations,
with the 3D torques being systematically weaker.
Within $\sim$1 $h_0$, the 2D and 3D profiles differ substantially
because of the different smoothing lengths employed (see \secref{sec:numerics}).}
\label{fig:tor_den}
\end{figure}

Here we offer an explanation for the origin of gap
streamers, sometimes called filaments, that have been 
observed in gap-opening simulations 
\citep[e.g.,][]{Devalborro06}.
These streamers arise from unsteady gap edges.
In our simulations, they are seen in both 2D
and 3D, but more so in 2D, and when $\Mp\geq 2\mj$.

The Rayleigh instability can operate at gap edges 
where the rotation profile is significantly modified
by gas pressure gradients.
This dynamical instability occurs wherever
the specific angular momentum decreases outward:
\begin{equation}\label{eqn:Rayleigh}
\frac{\partial l}{\partial R} < 0 \,,
\end{equation}
where $l=R^2 \Omega$ is the specific angular momentum.
Combining this with \eqnref{eqn:omega}
imposes a condition on the second radial derivative of the
gas pressure, and by extension the sharpness of gap 
edges \citep[e.g.,][]{Yang10}.
\cite{Kanagawa15a} have used this condition to model gap profiles.

\figref{fig:ang_momt} plots the azimuthally averaged 
values of $\partial l /\partial R$ for our 2D and 3D simulations.
For 3D, we plot the midplane values.
We find that when $\Mp\geq 2\mj$ in 2D, or $\Mp= 4\mj$ in 3D,
the gap edges are Rayleigh unstable.
These cases coincide with the appearance of unsteady gap edges
and enhanced streamers.
We have continued our 2D, $4\mj$ simulation to $10^4$ orbits to
verify that the gap edges remain unstable even at late times.

There is a slight difference between the 3D and 2D angular momentum
profiles, in that the former violates the Rayleigh criterion
to a lesser degree. Coinciding with this difference --- and arguably
causally related --- is the fact that the streamers leaking into the
gap are less prominent in 3D than in 2D (Figure \ref{fig:disks}).
In general, as demonstrated in Figure \ref{fig:tor_den},
the 3D torques are somewhat weaker than the 2D torques --- by up to
$\sim$50\% in the case of the $4\mj$ run.
We note that this implies that
a larger smoothing length $\rs$ in 2D may produce
a better 
match to 3D results for higher planet masses.

The Rayleigh instability may limit the extent to which gaps can be
emptied. As the explicit disk viscosity decreases, the gap
deepens, its edges become more prone to instability, and more
streamers leak into the gap (see also Section 4.2 of FSC14).
Thus the Rayleigh instability serves to provide an effective
viscosity, transporting angular momentum outward
to counter the planet's torque and establishing a 
``floor'' on the gap surface density.

\section{Summary and Discussion}
\label{sec:conclude}

Previous work on gap opening by planets has been plagued by
the question of whether 2D results carry over to realistic 3D disks.
We have concluded from our numerical study that at least some features do.
For our model parameters, gaps in 2D and 3D share nearly identical
depths and widths. Though this result could not have been predicted
without explicit 3D calculations like those we have undertaken,
the similarity between 2D and 3D is, in retrospect, sensible.
The 2D equations are merely the vertically integrated versions
of the 3D equations; in 3D, it must still be true that
the planetary and viscous torques, averaged over large enough volumes,
balance in steady state, as they do in 2D
(modulo the small residual from the global
viscous torque that eventually brings the entire disk onto the star).
The balance in 3D is not achieved by a static
cancellation of forces at every point in space,
but through a dynamic equilibrium between meridional
flows driven by the planet, and meridional flows
driven by viscosity.

To recapitulate our main findings:
\begin{enumerate}
\item Gaps carved in 3D by giant planets resemble their 2D counterparts
in depth and shape (Figures
\ref{fig:time}, \ref{fig:shape}, and \ref{fig:disks}).
\item Meridional flows are generated by the spatial
mismatch between the planet's torque and the disk's viscous
torque (Figures \ref{fig:flow_speed}, \ref{fig:flow_momt},
and \ref{fig:schm}). The flows extend vertically up to
a few scale heights, and radially over distances comparable
to the planet's orbital radius (\figref{fig:meri_speed}).
\item Velocities near a giant planet's orbit approach and
exceed the sound speed. Such fast flows may be used to distinguish
planetary gaps from gaps produced by other mechanisms
(\figref{fig:vz_map}).
\item The steepest gap edges become Rayleigh unstable
(\figref{fig:ang_momt}) and shed streamers that leak into the gap.
\end{enumerate}

There are some
caveats regarding our results. First, our viscous disk
models may not accurately represent protoplanetary disks, which 
may instead be turbulent \citep[e.g.,][]{Fromang11,Shi12}.
Second, limited by computational resources, we have not investigated 
how our results scale with disk viscosity or
temperature. A 
more thorough parameter space study should be possible
as computing technology continues to advance.
Third, insofar as accretion by the planet itself could further reduce
the ambient gas density,
the $\Sgap$ values found in this work should be considered upper limits.
Lastly, we have assumed midplane symmetry and only simulated the upper 
half of the disk. The validity of this assumption will need to be tested
by future work.

We close by discussing some directions for future work
on planet-disk interactions.

\subsection{Planet Accretion}

Disk flows are directed onto the planet above its poles, and away from
the planet in the equator plane
\citep{Kley01,Klahr06,Tanigawa12,Szulagyi14,Morbidelli14,Ormel15b,Fung15}.
The giant planets simulated here should be in the final ``runaway'' phase
of their mass accretion history, when the self-gravity of their gas envelopes is significant. 
Accretion rates for planets in this regime, such as those simulated by \citet{DAngelo2003} and \citet{Machida10}, and calculated analytically by \citet{Tanigawa16}, show a clear dependence on the ambient gas density. The gap depth should therefore matter for planet accretion rates and luminosities.
The converse should also be true. To assess the effect of
planetary accretion on the gap density, we would need to adopt a different
planet boundary condition than our present smoothing length prescription.

\subsection{Meridional Flows}
The large-scale vertical and radial flows induced by giant planets
may affect a number of disk processes: the radial transport of grains
(see, e.g., \citealt{Ogliore09} for possible implications of the
{\it Stardust} mission); dust filtration at gap edges which can
impact the appearance of transitional disks \citep[e.g.,][]{Zhu14};
the ability of dust to settle vertically, 
which can affect the disk's spectral energy distribution
\citep[e.g.,][]{Chiang01};
and disk chemistry
\citep[e.g.,][]{Bergin07,Dutrey14}.

\subsection{Gap Opening by Multiple Planets}
The wide radial extents of the cavities in transitional disks have led
some to invoke gap opening by multiple planets \citep{Zhu11}.
Simulations in 2D of multi-planet gaps have found that common gaps
(the merged gaps of more than one planet) have significantly higher gas density
than single-planet gaps \citep{Duffell15},
making it more difficult to reproduce the transparency
of transition disk holes. We plan to test this result in 3D
in a forthcoming paper.

\acknowledgments 
We thank Willy Kley, Fr\'{e}d\'{e}ric Masset, Alessandro Morbidelli, 
Ji-Ming Shi, and Judit Szul\'{a}gyi for helpful exchanges.
JF gratefully acknowledges support from 
the Natural Sciences and Engineering Research Council of 
Canada, and the Center for Integrative Planetary Science 
at the University of California, Berkeley. EC acknowledges 
support from grant AST-1411954 awarded by the 
National Science Foundation, and NASA Origins grant 
NNX13AI57G.

\bibliographystyle{apj}
\bibliography{Lit}

\end{document}